\documentstyle[twoside,fleqn,espcrc2,epsfig]{article}

\title{Potential between external monopole and antimonopole in SU(2) lattice gluodynamics
\thanks{Presented by Ch.~Hoelbling. This research was supported
in part under DOE grant DE-FG02-91ER40676 and by the
U.S.~Civilian Research and Development Foundation for
Independent States of FSU (CRDF) award RP1-187.}}

\author{Ch.~Hoelbling$^{\rm a}$,
C.~Rebbi\address{Boston University Physics
    Department\\
    590 Commonwealth Avenue\\
    Boston MA 02215, USA}
        and
V.~A.~Rubakov\address{Institute for Nuclear Research
of the Russian Academy of
  Sciences\\
60th October Anniversary Prospect 7a\\
Moscow 117312, Russian Federation}
}

\begin{document}
\newcommand{\bp}{{\beta^\prime}}
\newcommand{\Tr}{\mbox{Tr}}

\begin{abstract}

 We present the results of a study of the free energy of a monopole pair in pure SU(2) theory at finite temperature, both below and above the deconfinement transition. We find a Yukawa potential between monopoles in both phases. At low temperature, the screening mass is compatible with the lightest glueball mass. At high temperature, we observe an increased screening mass with no apparent discontinuity at the phase transition.

\end{abstract}

\maketitle

\section{Introduction}

 Since it was suggested by 't~Hooft \cite{tH76} and Mandelstam \cite{Ma76} that monopole condensation in nonabelian gauge theories may be an explanation for the confinement mechanism, many studies have been devoted to the magnetic properties of these theories \cite{KT98,Co98,En98,DGT82,BLS81}.

 In the present study, we probe the vacuum structure of pure SU(2) gauge theory by inserting a static monopole-antimonopole pair into the vacuum and measuring its free energy at different separations and different temperatures. While in the classical theory there is a Coulomb potential between SU(2) monopoles, we expect the quantum theory to show a Yukawa-potential if there is a monopole condensate, and a Coulomb-potential, if quantum fluctuations do not produce a magnetically screening object.

\section{Simulation method}

\begin{figure}
\begin{center}
\epsfig{file=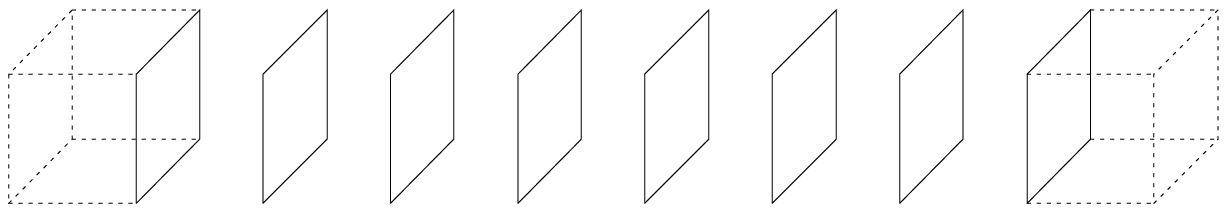,width=7cm}
\end{center}
\caption{\label{chain}Monopoles sit at the end of a magnetic flux tube}
\end{figure}

 The method we use for inserting a static SU(2) monopole pair on the lattice was put forward by 't~Hooft and others \cite{tH78,MP79,MP80,Ya79,UWG80,SS81} and essentially amounts to introducing a magnetic flux tube by multiplying the couplings with an element of the center of the gauge group ($\beta\rightarrow -\beta$ for SU(2)) on a string of plaquettes in every timeslice (schematically shown in fig.\ref{chain}).

 This tube carries a flux of $\Phi=\pi\sqrt{\beta}$, so at each end of the flux tube is a elementary cube with net magnetic outflux $\Phi=\pm\pi\sqrt{\beta}$. But since for a compact lattice theory magnetic flux is only defined modulo $2\pi\sqrt{\beta}$, both monopoles are indistinguishable and coincide with their own antimonopole.

 In order to measure the free energy of a monopole pair, we split the SU(2) action into two parts with different couplings
\begin{equation}
\label{act}
S(\beta,\bp)=\frac{1}{2}\left(\beta\sum_{P\not\in M}\Tr(U_P)+\bp\sum_{P\in M}\Tr(U_P)\right)
\end{equation}
 where $M$ is the set of all the plaquettes in the monopole string. In the case where $\bp=\beta$, (\ref{act}) is the action of a plain SU(2) theory and for $\bp=-\beta$ it is the action of SU(2) with the static monopole pair. So the free energy difference upon a monopole pair insertion is
\begin{equation}
\label{freen}
\Delta F=-T\ln\frac{Z(\beta,-\beta)}{Z(\beta,\beta)}
\end{equation}
 where $T=1/N_t a$ is the temperature of the system and $Z(\beta,\bp)$ is the partition function
\begin{equation}
Z(\beta,\bp)=\sum_{\mathcal C} e^{-S(\beta,\bp)}
\end{equation}

 We can calculate $\Delta F$ using the Ferrenberg-Swendsen multihistogram method \cite{FS} on histograms of the flux tube energy
\begin{equation}
E^{\prime}=\frac{1}{2}\langle\sum_{P\in M}\Tr(U_P)\rangle_\bp
\end{equation}
for a set of $\bp$'s between $-\beta$ and $\beta$ (where the separation between successive $\bp$'s has to be small enough, that neighboring histograms have a substantial overlap).

 Our simulations were performed with couplings between $\beta=2.476$ and $\beta=2.82$ on systems of volumes, ranging from $N_x\times N_y\times N_z=16^2\times 32$ to $N_x\times N_y\times N_z=32^2\times 64$. The time extent in our simulation varied between $N_t=2$ and $N_t=16$. For each system, we measured the free energy at monopole separations from $a$ to $6a$.

 We used a combined overrelaxation and 3-hit Metropolis update algorithm. For every simulation, we started at $\bp=\beta$ with $5000$ thermalization updates followed by $200$ to $800$ independent measurements of $E^{\prime}$, separated by $50$ updates. We then decreased $\bp$ in $11$ to $61$ steps to $\bp=-\beta$ and performed $500$ thermalization updates followed again by the measurement updates. Here, a single $E^{\prime}$ measurement consisted of an average over $384$ configurations of the plaquettes on the monopole string on a fixed background configuration.

\section{Results}

\begin{figure}
\epsfig{file=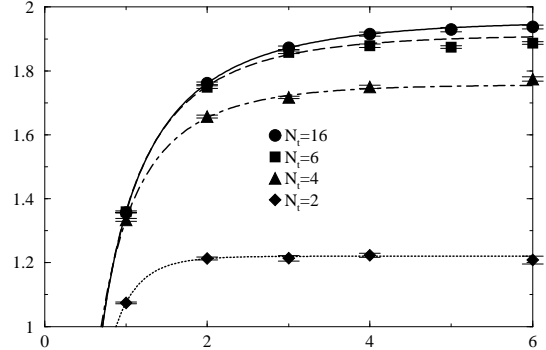,width=7cm}
\caption{\label{fmass}Free energy of the monopole-antimonopole pair vs.~distance in lattice units at $\beta=2.6$ for different values of $N_t$. The lines represent fits to a Yukawa potential.}
\end{figure}

 In fig. \ref{fmass} we plot the free energy vs.~monopole pair separation for a $N_x\times N_y\times N_z=20^2\times 40$ system at $\beta=2.6$ and different temperatures. The lines are fits to a Yukawa-potential
\begin{equation}
F(r)=F_0-c\frac{e^{-mr}}{r}
\end{equation}

\begin{table}
\begin{tabular}{r|l|l|l|l}
$N_t$ & $T/\sqrt{\sigma}$ & $m/\sqrt{\sigma}$ & $ma$ & Q\\
\hline
2 & 3.676 & 17.3(4.1) & 2.29(55) & 0.41 \\
4 & 1.838 & 5.43(59) & 0.720(78) & 0.03 \\
6 & 1.225 & 4.13(41) & 0.548(54) & 0.02 \\
16 & 0.460 & 3.24(42) & 0.430(56) & 0.96
\end{tabular}
\caption{\label{masstab}Screening masses for $\beta=2.6$ in units of the string tension at zero temperature $\sqrt{\sigma}$ and in lattice units. For comparison, the temperature of the system is given. Q is the fit quality.}
\end{table}

\begin{table}
\begin{tabular}{r|l|l|l}
$\beta$ & $m_g a$ & $ma$ & $m_ra$ \\
\hline
2.5 & 0.66(2) & 0.46(8) & 0.64(20)  \\
2.6 & 0.51(3) & 0.43(6) & 0.53(12) 
\end{tabular}
\caption{\label{gb}Comparison of the screening masses ($m,m_r$) in the confined phase to the lightest glueball mass ($m_g$, taken from \cite{E98}). $m_r$ is the screening mass from a restricted fit ignoring the data point at monopole separation $a$.}
\end{table}

 Table \ref{masstab} shows the screening masses obtained from this fit. A Coulombic behavior ($m=0$) is clearly ruled out. The quality of a Coulomb fit is $Q\approx 10^{-7}$ for $N_t=16$ and even worse ($Q<10^{-15}$) for all other cases.

 At low temperature, we can compare our screening masses to the known mass of the lightest glueball state (table \ref{gb}). The masses are roughly in agreement, especially when one ignores the data point at separation $a$, which is most affected by discretization errors.

\begin{figure}
\epsfig{file=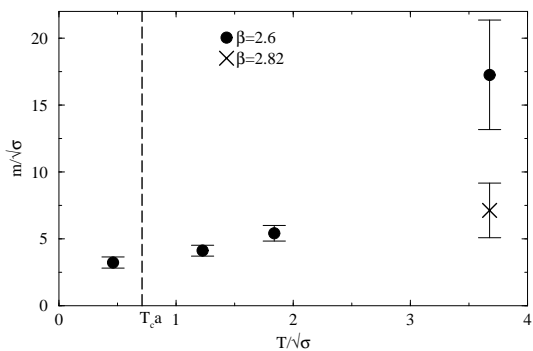,width=7cm}
\caption{\label{fhit} Plot of the screening mass vs.~temperature. The last data point for $\beta=2.6$ is from a simulation with $N_t=2$ and should be taken with caution. The object acquires a thermal mass and there is no indication for the deconfinement transition in this observable. The dashed line indicates the critical temperature (taken from \cite{FHK92}).}
\end{figure}

 In fig. \ref{fhit} we plot the screening masses vs.~temperature. There clearly is an increase of the screening mass with temperature, but we can see no signal of the phase transition.


\section{Conclusion}

 We have studied the free energy of a monopole pair in pure SU(2) gauge theory at finite temperature. We find, that in both phases it exhibits a screened behavior. At low temperature, the screening mass is roughly in agreement with the mass of the lightest glueball state. At high temperature, we observe an increase in the screening mass with no apparent discontinuity at the phase transition.

\end{document}